# The Superconducting Magnet for ADS Injection-I

Xiangchen Yang[1], Da Cheng, Quanling Peng[1], Yuan Chen[1], Fengyu Xu[2], Anbin Chen[2]

1) Institute of High Energy Physics, Chinese Academy of Sciences, Beijing 100049, China
2) Harbin Institute of Technology, Heilongjiang, 150006, China

**Abstract**

The cryomodule I (CM1) for ADS Injection-I had been designed, fabricated and online tested. The CM1 contains seven superconducting magnets, seven superconducting spoke cavities and seven beam position monitors. The superconducting magnet, which is a kind of multifunction magnet in a 170mm length axial space, contains a solenoid for beam focusing and two correctors for orbit correction. The design goals for the magnets are to meet the required integral field strength and to reduce the leakage field of less than 2 G at the nearby superconducting spoke cavities. The 2.1K, 31 mbar cryogenic system, where the magnets and spoke cavities shared with, force the magnet must select a kind of conduction cooled current leads. The first one of the batch magnets was tested in a vertical Dewar in HIT in July, 2014, the measurement results met the design requirements. Online operation of CM1 in September, 2015 showed that seven magnets can work at 230A under 4.2K and 2.1K respectively. This paper will present the magnet and current lead design, field measurement and the installation progress.

## 1. Overview of ADS superconducting magnet

The Accelerator Driven Sub-critical System (ADS) containing two cryomodules in injection-I is to realize the safe disposal of nuclear waste based on 2.1K, 31 mbar cryogenic system [1]. Figure 1 shows the Schematic layout of ADS injection I [2]. Two cryomodules, each consisting of 7 superconducting spoke cavities, 7 superconducting magnets and 7 beam position monitors, are used to accelerate the proton beam from 3.2 MeV to 10 MeV. Each superconducting magnet package consists of three function magnets with a solenoid for beam focusing, a horizontal dipole corrector (HDC) and a vertical dipole corrector (VDC) for beam orbit correction. The HDC and VDC dipole coils are installed at the bore of the solenoid coils. The required focusing field strength of the solenoid is 1.10 $T^2$-m, where the maximum integrated field strength for HDC and VDC is $1.6 \times 10^{-3}$ T-m. Because of the negative pressure environment, helium gas can't flow from the cold environment to the warm end to cool the current leads, so conduction cooled current leads was selected.

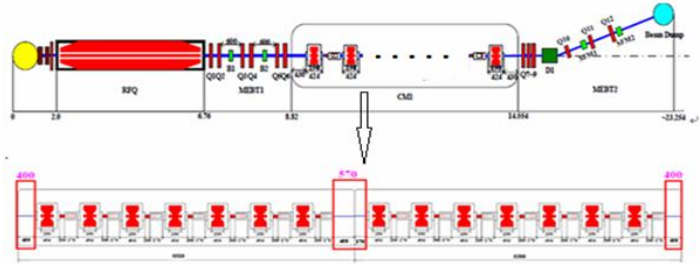

Figure 1 Schematic layout of ADS injection I. Two cryomoudles for the accelerator components installation are shown in below.

## 2. Physical and mechanical design of the magnet cryostat

In order to install the magnet and the spoke cavities into the new compacted cryomoudle, the length of the superconducting magnet was shortened from 300 mm to 170 mm. Detailed physical and mechanical design of the 300 mm long magnet can be shown in reference [3]. The short magnet was realized by removing two conflat flanges, and replaces them with the compress flanges that with the connection screw holes. Another modification was shortened main solenoid and two bucking solenoids somehow to realize a compact magnet design. Two aluminum rings were used as the vacuum seals to connect magnet cryostat with the upstream spoke cavity and the downstream BPM.

Figure 2 shows the 2D case of the solenoid field, axial symmetry is used here for the magnetic field calculation. The solenoid field is realized by a main solenoid coil, two bucking solenoid coils, the iron yoke, on the other hand, is used to confine the magnetic flux and to reduce the leakage field in further more [4]. With the field optimization, the leakage solenoid field at 270mm away from the solenoid center is reduced less than 2G. The axial repulsive force on the bucking solenoid is 44 kN at the design current of 210 A, which must be eliminated by pre-press forces during the magnet assembly.

The winding coils for the two correctors, HDC and VDC, are saddle shaped at the inner surface of the main solenoid coil. The reason for doing such a selection lies in that to save space, and also to reduce the total magnetic energy stored. HDC and VDC coils were first wound in flat pattern, and then wrapped around the support tube to create the saddle shape, and then fixed them on by using Stycast 2850 FT epoxy.

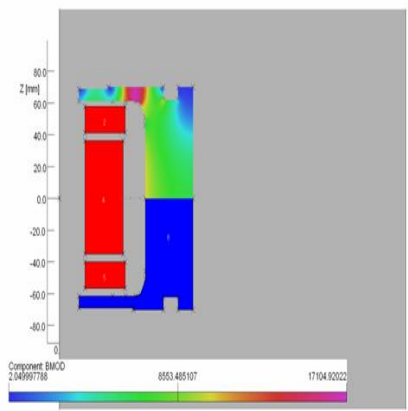

Figure 2. Physical design of the superconducting solenoids. Dark red region as shown in the front return yoke is beyond from saturation.

Figure 3 shows the longitunal cross section of the magnet cryostat. The support tube, where the corrector coils directly wound on, also serves as the beam vacuum chamber and as the inner helium vessel. The compress flange at both sides of magnet cryostat provides pre-press forces and realizes the sealing with the spoke cavity and the BPM. During the magnet excitation, the large repel force from the bucking solenoids will then directly transfer to the compress flange, the weld joint where the compress flange and support tube welded on was optimized to withstand that force.

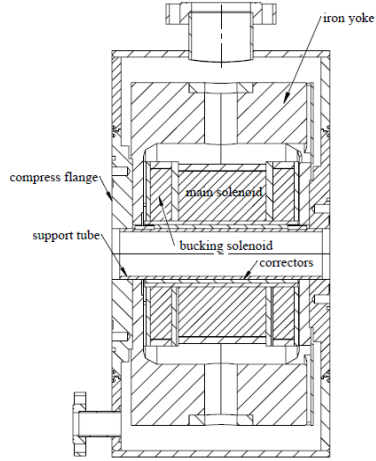

Fig. 3. Schematic overview of the magnet cryostat

**3. Design of the current lead**

Current leads are of great importance as they are applied to deliver room-temperature electrical power to the cryogenic environment. Liquid helium at 30 mbar and 2.1 K is to be used as the coolant for the superconducting spoke cavity. To simplify the cryogenic system, the superconducting magnets use the same state of liquid helium as that of the spoke cavity. Since the liquid helium is at low pressure, the evaporated helium cannot be returned to the cryogenic system. Consequently, the conducted current leads were used for the magnets. Heat generated in the current leads will be removed by thermo anchors, similar to those of the LHC corrector magnets [5][6]. For the ADS superconducting magnet, two-stage thermo anchors are used, with one connected to the liquid nitrogen shield (77 K), and the other to the 5K helium gas shield. Three pairs of current leads are needed for each magnet, with one pair carrying 210 A for the solenoid and two pairs carrying 15 A for the correctors [7]. The length of each section was optimized according to the cross section of TP2 copper (RRR=20) and the 210 A design current. The optimization length of the three separated sections, which is respectively from 300K to 80K, from 80K to 5K and from 5K to 2K, can be expressed as:

$$\frac{L}{A} = \frac{1}{I}\int_{T_0}^{T_1} \frac{\lambda(T)}{\sqrt{L_0(T_1^2 - T^2)}} dT \qquad (1)$$

Where L is the optimized length, A is the cross area of the current leads, $I$ is the design current, $T_1$ is the temperature at warm end, $\lambda(T)$ is the thermal conductivity of leads as the function of the temperature for each section, $L_0 = 2.45 \times 10^{-8}$ W Ω K$^{-2}$ is the Franz constant.

With that of length optimization, the minimum at each section is:

$$(Q_0)_{min} = I\sqrt{L_0(T_1^2 - T_0^2)} \qquad (2)$$

Pure copper coated on the brass bar as the current leads shown in reference [5][6] was finally abandoned since the complex process and the expensive price. Fig. 4 shows the 3D model of the whole magnet assembly. The leads contain four hollow tubes, one is used for holding signal lines, and other two tubes with each one carry a current lead for solenoid coil, whereas the last one contains four copper bars for the two corrector coils. Electrically insulated between the current leads and the outer stainless steel tube is realized by two layers of kapton tubes. Considering the installation process, the current leads need to be temporarily compressed, so it is designed to a kind of spring. As shown in Fig. 4, three layers of thermal anchor clamps the four stainless tubes together by screws, they are connected to the thermal anchor seat by the copper braids. The thermal anchor seats are hard soldered onto the cryogenic pipelines which are welded on the cold shield.

The gas collected tank, where the helium gas contained inside, also severs as interface for the current leads and the power supply. The seat of the gas tank is welded onto the surface of the cryomoudle.

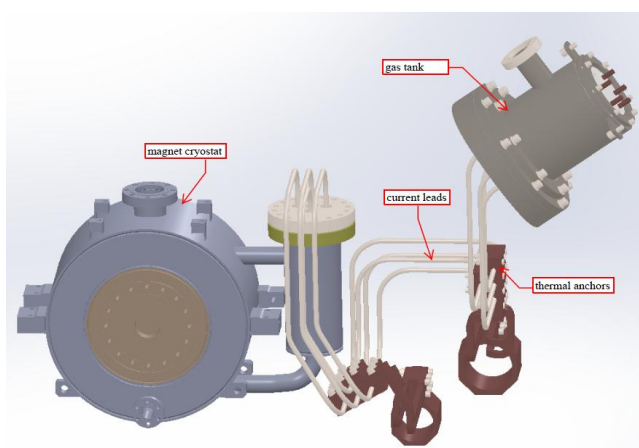

Fig 4. 3D engineering drawing of the magnet system, at left is the magnet cryostat, four bending tubes are the current leads, at up right is the gas tank.

**4. Field measurement**

A moveable field measurement platform will be seated on the top flange. The bare superconducting magnet, where the outer helium vessel is not welded on, will be fixed vertically on the hoisting flange at the bottom. An isolation tube is deliberately designed for the magnetic field measurements, it consists of two concentric 316L stainless steel tubes separated by a vacuum jacket. It provides thermal isolation from the liquid helium to the air. The superconducting magnet is hanged on the hoisting flange vertically which can keep the beam vacuum tube coaxial with the isolation tube. During the field measurements, the movable platform will drive the three-axis Hall probe move vertically inside the isolation tube in each 2mm step in a maximum distance of 460 mm [8].

The designed integral field maximum for the solenoid is 0.4 T-m, for HDC or VDC, it is 1600 G-cm. The vertical test was preceded in HIT in July. 2014. The magnet training process indicates that the solenoid magnet can finally reach to 260 A for 15 minutes after several times of natural quench. Field measurement tests were carried on at several excited currents as 100A, 150A, 190A and 230A. The leakage field is well limited to less than 2G at the distance of 270 mm away from the solenoid center under 230A, which satisfies the requirements at the upstream and downstream spoke cavities. Fig. 5 shows the field distribution generated by different excited current.

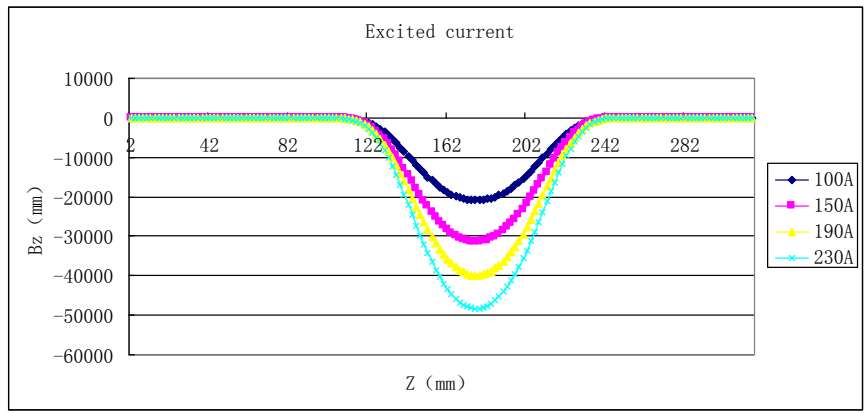

Fig. 5 Magnetic field distribution under different excited current

**5. Vacuum check of the whole magnet assembly**

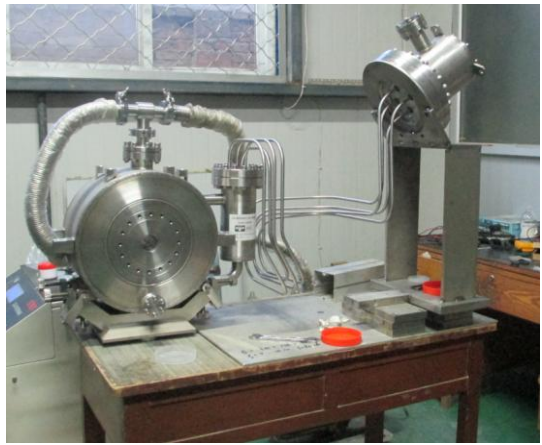

Fig. 6 Vacuum check of the whole magnet assembly

Vacuum check of the whole magnet assembly was performed before it is shipped to the next installation process. Magnet cryostat and the current leads are separately passed through the high pressure test and the vacuum check. The whole magnet assembly need more test such as cold shock test, high pressure $N_2$ gas filling test, and then the vacuum check. The pressure for the high pressure test is 6 kg/cm$^2$, leakage rate must be less than $1 \times 10^{-9}$ Pa.m$^3$/s. Figure 6 shows the picture of vacuum check of the whole magnet assembly.

**6. Magnet installation process.**

The superconducting magnets installation was completed in IHEP in July, 2015. The installation processes are shown as the followings. First, the beam aperture for each of the magnet was cleaned by ultrasonic wave. The magnets were connected to the other accelerator components like BPM and spoke cavities to form a chain of superconducting components in a high clean room. The chain of superconducting components will be moved out of the clean room and sit on the support platform of the CM1. Current leads connection for the copper bars and the superconducting cables, signal wires connection were completed by soldering on a temporary table. Figure 7 shows the current leads soldering process.

At last, the chain of superconducting components with the cold shield covered was push into the body of the cryomoudle. The gas tank, which is part of the superconducting magnet was then installed onto the surface of the cryomoudle. In order to satisfy the sealing and the removable requirements, there are five seal interfaces in total for the gas tank. Fig. 9 shows the picture of the moment when the installation was completed. High voltage test to the ground is necessary for each installation process.

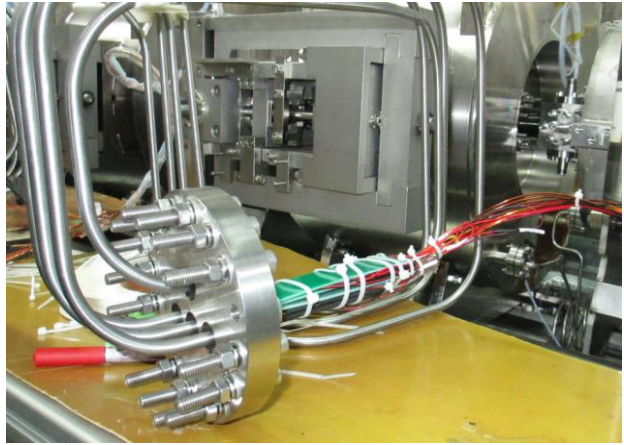

Fig. 7. The current lead soldering process.

When the leak detection under normal temperature and liquid nitrogen temperature was finished, CM1 was then moved to the tunnel and connected to the upstream and downstream accelerator components. Online cryogenics test show that, superconducting magnets in CM1 can operate normally in September, 2015. They all can be operated at 4.2K and 2.1K respectively in 230 A for 15 minutes without quench.

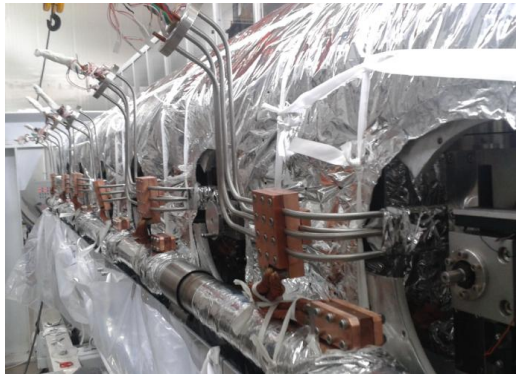

Fig. 8. Thermal anchors were connected to the anchor seats during the magnet assembly process, also as can be shown; the multilayer insulation was covered on the 5K cold shield.

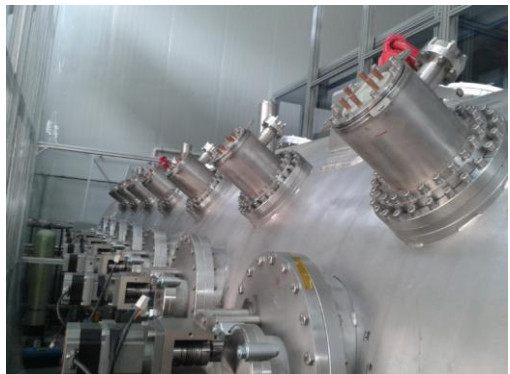

Fig. 9. Gas tank installation for the magnet

**7. Conclusion**

As difficult as it was, the design and manufacture of the seven superconducting magnets and current leads of CM1 was still successfully realized. And to make things even better, the field meets the design demands and the online test shows that the magnets can work in a rather stable state. In the near future, the batch production of another seven magnets would be installed in the subsequent cryomodule II. Although the whole process of design, manufacture and installation was accompanied with many problems on account of the complexity of equipment technology and the technical difficulty. It also accumulated a wealth of experience for the future work.